\documentclass[10pt,twocolumn,english,aps,prl]{revtex4-1}
\usepackage[T1]{fontenc}
\usepackage[latin9]{inputenc}
\usepackage{color}
\usepackage{babel}
\usepackage{float}
\usepackage{amssymb}
\usepackage{graphicx}
\usepackage[unicode=true,pdfusetitle,
 bookmarks=true,bookmarksnumbered=false,bookmarksopen=false,
 breaklinks=true,pdfborder={0 0 1},backref=false,colorlinks=true]
 {hyperref}

\makeatletter
\usepackage{pdfpages}
\usepackage{pgffor}
\usepackage[caption=false]{subfig}
\AtBeginDocument{\let\LS@rot\@undefined}

\@ifundefined{showcaptionsetup}{}{%
 \PassOptionsToPackage{caption=false}{subfig}}
\usepackage{subfig}
\makeatother

\begin{document}

\title{A synthetic gauge field for two-dimensional time-multiplexed quantum
random walks}

\author{Hamidreza Chalabi}
\email{hchalabi@umd.edu}

\selectlanguage{english}%

\affiliation{Department of Electrical and Computer Engineering, The University
of Maryland at College Park, College Park, MD 20742, USA}

\author{Sabyasachi Barik}
\email{sbarik@umd.edu}

\selectlanguage{english}%

\affiliation{Department of Electrical and Computer Engineering, The University
of Maryland at College Park, College Park, MD 20742, USA}

\author{Sunil Mittal}
\email{smittal@umd.edu}

\selectlanguage{english}%

\affiliation{Department of Electrical and Computer Engineering, The University
of Maryland at College Park, College Park, MD 20742, USA}

\author{Thomas E. Murphy}
\email{tem@umd.edu }

\selectlanguage{english}%

\affiliation{Department of Electrical and Computer Engineering, The University
of Maryland at College Park, College Park, MD 20742, USA}

\author{Mohammad Hafezi}
\email{hafezi@umd.edu}

\selectlanguage{english}%

\affiliation{Department of Electrical and Computer Engineering, The University
of Maryland at College Park, College Park, MD 20742, USA}

\author{Edo Waks}
\email{edowaks@umd.edu}

\selectlanguage{english}%

\affiliation{Department of Electrical and Computer Engineering, The University
of Maryland at College Park, College Park, MD 20742, USA}
\begin{abstract}
Temporal multiplexing provides an efficient and scalable approach
to realize a quantum random walk with photons that can exhibit topological
properties. But two dimensional time-multiplexed topological quantum
walks studied so far have relied on generalizations of the Su-Shreiffer-Heeger
(SSH) model with no synthetic gauge field. In this work, we demonstrate
a 2D topological quantum random walk where the non-trivial topology
is due to the presence of a synthetic gauge field. We show that the
synthetic gauge field leads to the appearance of multiple bandgaps
and consequently, a spatial confinement of the random walk distribution.
Moreover, we demonstrate topological edge states at an interface between
domains with opposite synthetic fields. Our results expand the range
of Hamiltonians that can be simulated using photonic random walks.
\end{abstract}
\maketitle
Photonics provides a compelling platform to study quantum random walks
\cite{Aharonov1993}. Photons can propagate over long distances without
losing coherence, enabling complex random walks that can implement
various quantum computing algorithms \cite{Kempe2003,Shenvi2003,Childs2003},
and also simulate a broad range of quantum Hamiltonians \cite{Childs2013}.
Photonic random walks in both one and two dimensions can be implemented
in spatial degrees of freedom using beam splitters \cite{DeNicola2014,Sansoni2012,Broome2010}
or coupled waveguide arrays \cite{Poulios2014,Tang2018,Tang2018a}.
But such approaches are difficult to scale to large number of steps,
particularly when going to higher dimensions. 

Synthetic spaces provide an alternative approach to scale the state-space
of the walker without requiring complex photonic circuits. Examples
of synthetic spaces include frequency \cite{Navarrete-Benlloch2007,Bouwmeester1999,Lin2016,Lin2018,Yuan2018},
orbital angular momentum \cite{Goyal2013,Cardano2015,Cardano2016,Cardano2017},
and transverse spatial modes as recently experimentally realized \cite{EranLustigSteffenWeimannYonatanPlotnikYaakovLumerMiguelA.BandresAlexanderSzameit2018}.
Time-multiplexing is another synthetic space that is particularly
easy to work with \cite{Schreiber2010,Schreiber2011,Schreiber2012,Nitsche2016,Barkhofen2017,Chen2018}.
Time-multiplexed random walks have the advantage that they can span
an extremely large state-space with only a few optical elements and
can efficiently scale to higher number of walker dimensions. 

Recently, time-multiplexed quantum walks have been used to explore
topological physics and the associated edge states in both one and
two dimensional systems \cite{Barkhofen2017,Chen2018}. Most realizations
of such topological quantum walks are based on the split-step quantum
walk protocol \cite{Kitagawa2010,Kitagawa2010a,Kitagawa2012,Kitagawa2012a}.
Similar to the Su-Shreiffer-Heeger (SSH) model, here the non-trivial
topology is a result of the direction-dependent hopping strength between
the lattice sites. However, many of the most interesting topological
Hamiltonians, such as the integer quantum Hall effect \cite{Klitzing1980},
the Haldane model \cite{Haldane1988}, and the quantum-spin Hall effect
\cite{Konig2007}, require gauge fields that generate direction-dependent
hopping phases. So far, time-multiplexed random walks with synthetic
gauge fields have only been implemented in 1D, which severely restricts
the number of topological Hamiltonians that can be explored. 

Here, we demonstrate a two-dimensional topological synthetic gauge
field in a time-multiplexed quantum walk. We show that in our discrete-time
quantum walk, the pseudo-energy band structure exhibits multiple bandgaps
depending on the magnitude of the synthetic gauge field. These bandgaps
result in the confinement of the random walker, as opposed to ballistic
diffusion that would otherwise occur \cite{Yalcnkaya2015}. Moreover,
we demonstrate the presence of multiple topological edge bands at
an interface between two domains with opposite magnetic fields. Because
of the presence of two topological edge bands, our system supports
two sets of non-degenerate topological edge states that travel in
forward and backward directions along the interface. 

\begin{figure}
\subfloat[\label{fig:Fig1a}]{

}\subfloat[\label{fig:Fig1b}]{}\includegraphics{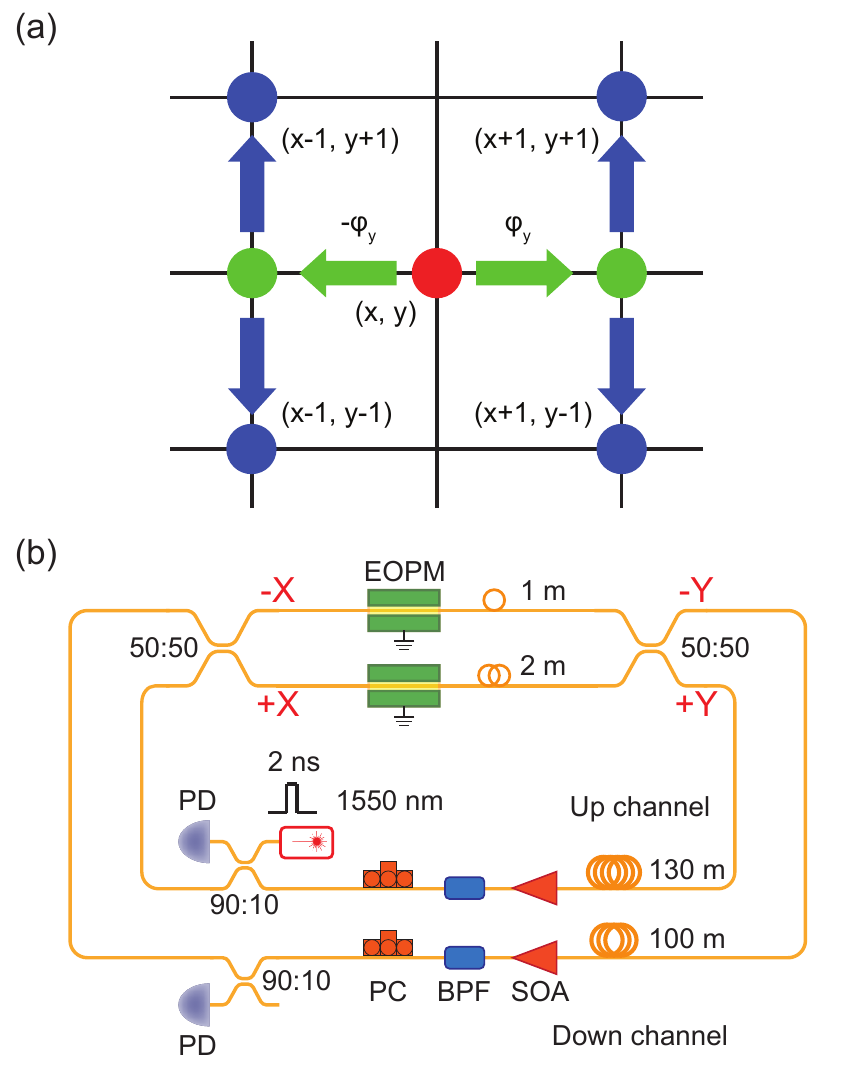}

\caption{\label{Fig1}(a) Schematic explaining the possible movements of a
walker at spatial position $\left(x,y\right)$ along with applied
phase shifts during each step of the random walk. (b) Schematic of
the 2D random walk setup describing the details of the experimental
setup. PD: photodetector, BPF: band pass filter, SOA: semiconductor
optical amplifier, EOPM: electro-optic phase modulator, and PC: polarization
controller.}
\end{figure}

To implement a gauge field, pulses must accumulate a net phase shift
when walking around a closed trajectory. Figure \ref{fig:Fig1a} shows
how we implement this condition. We apply  a phase shift of $y\phi$
when the walker moves to the right, and $-y\phi$ when the walker
steps to the left, where $y$ is the vertical coordinate of the walker.
This phase convention realizes a uniform magnetic field in the Landau
gauge. Consequently, this phase modulation scheme can be harnessed
as an artificial gauge field affecting the evolution of the random
walk. 

In our time-multiplexed photonic random walk, optical delays map the
walker state-space into time-delays of optical pulses. Similar to
earlier studies \cite{Schreiber2012,Chen2018} of two dimensional
quantum random walks, we implement these delays using a pair of nested
fiber delay loops. Figure \ref{fig:Fig1b} shows the schematic of
the experimental setup, and the full details are explained in the
Supplementary Material \cite{supplem}. The experimental setup essentially
consists of two beam-splitters with their ports connected to fibers
of different lengths such that they map the $\pm x$ and $\pm y$
directions to different time delays. One complete propagation of an
optical pulse around the loop is then equivalent to hopping of the
walker to one of the four possible corners in the synthetic space
(Fig. \ref{fig:Fig1a}). Two semiconductor optical amplifiers (SOAs)
are employed to partially compensate for the losses that the optical
pulses experience in each round trip. In this setup, we study the
random walk distribution at each time step via two photodetectors
analyzing two channels that we refer to as the up and down channels
as labeled in Fig. \ref{fig:Fig1b}. We initialize the random walk
through a single incident laser pulse that is injected into the up
channel starting the evolution of random walk distribution from the
origin in synthetic space. Here, we have analyzed the evolution of
the random walk based on the pulses detected in the down channel.
The electro-optic modulators which are driven by programmable voltage
waveforms are used for producing the desired phase shifts to generate
the synthetic gauge field.

\begin{figure}
\subfloat[\label{fig:Fig2a}]{}\subfloat[\label{fig:Fig2b}]{}\subfloat[\label{fig:Fig2c}]{}\subfloat[\label{fig:Fig2d}]{}\includegraphics{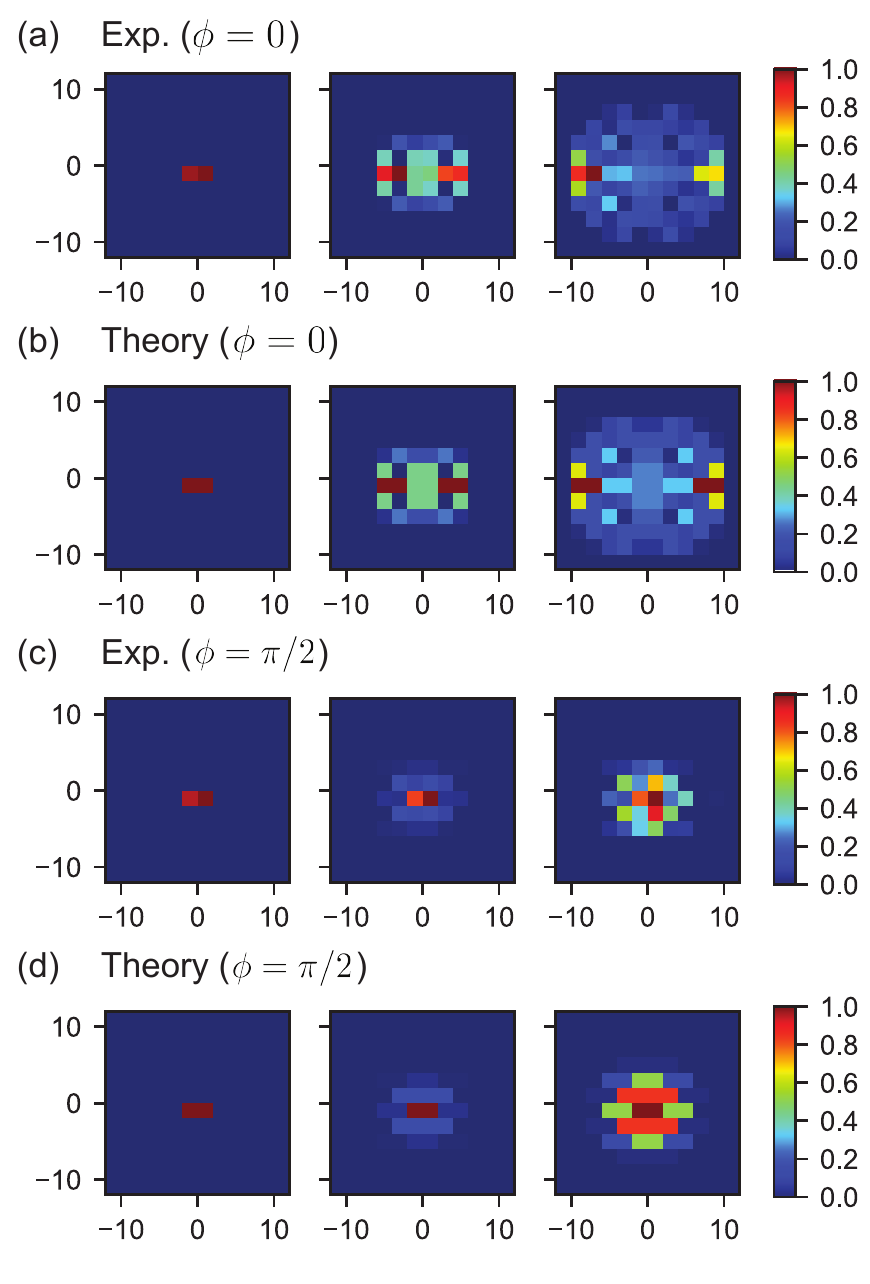}\caption{\label{Fig2}(a) Experimental observations and (b) theoretical predictions
of the evolution of the random walk distribution under no phase modulation.
(c) Experimental observations and (d) theoretical predictions of the
evolution of the random walk distribution under linearly dependent
phase modulation $\phi_{y}=y\phi$ for the case of $\phi=\pi/2$.
The left, middle, and right columns show the distributions at time
steps of 1, 5, and 9, respectively. In these plots all the distributions
are normalized to their maximum. }
\end{figure}

Figure \ref{Fig2} compares the evolution of the random walk distribution
with and without an applied gauge field. Figure \ref{fig:Fig2a} shows
the experimental results for the evolution under no applied gauge
field. In this figure, the distribution of the random walker is shown
at three different time steps of 1, 5, and 9. In the absence of a
gauge field, the random walk exhibits rapid diffusion. These results
are consistent with the theoretical predictions shown in Fig. \ref{fig:Fig2b}.
Figure \ref{fig:Fig2c} shows the experimental results for the evolution
of the random walk distribution in the presence of a gauge field with
$\phi=\pi/2$. The gauge field leads to suppressed diffusion and confinement
of the random walk distribution. The experimental results shown in
Fig. \ref{fig:Fig2c} for the case of a gauge field with $\phi=\pi/2$
are consistent with the theoretical predictions shown in Fig. \ref{fig:Fig2d}. 

We calculate the fidelity of the measured distributions relative to
the theoretical distributions ($P_{th}$) based on $F\left(n\right)=\sum_{x,y}\sqrt{P_{th}\left(x,y;n\right)P_{exp}\left(x,y;n\right)}$.
For the case of no applied gauge field, we obtained fidelities of
0.999, 0.996, and 0.985 for time steps of 1, 5, and 9, respectively.
Similarly, for the random walk under the gauge field we determined
fidelities of 0.999, 0.993, and 0.972 for time steps of 1, 5, and
9, respectively. 

\onecolumngrid 

\begin{figure}[H]
\subfloat[\label{fig:Fig3a}]{}\subfloat[\label{fig:Fig3b}]{}\subfloat[\label{fig:Fig3c}]{}\subfloat[\label{fig:Fig3d}]{}\includegraphics{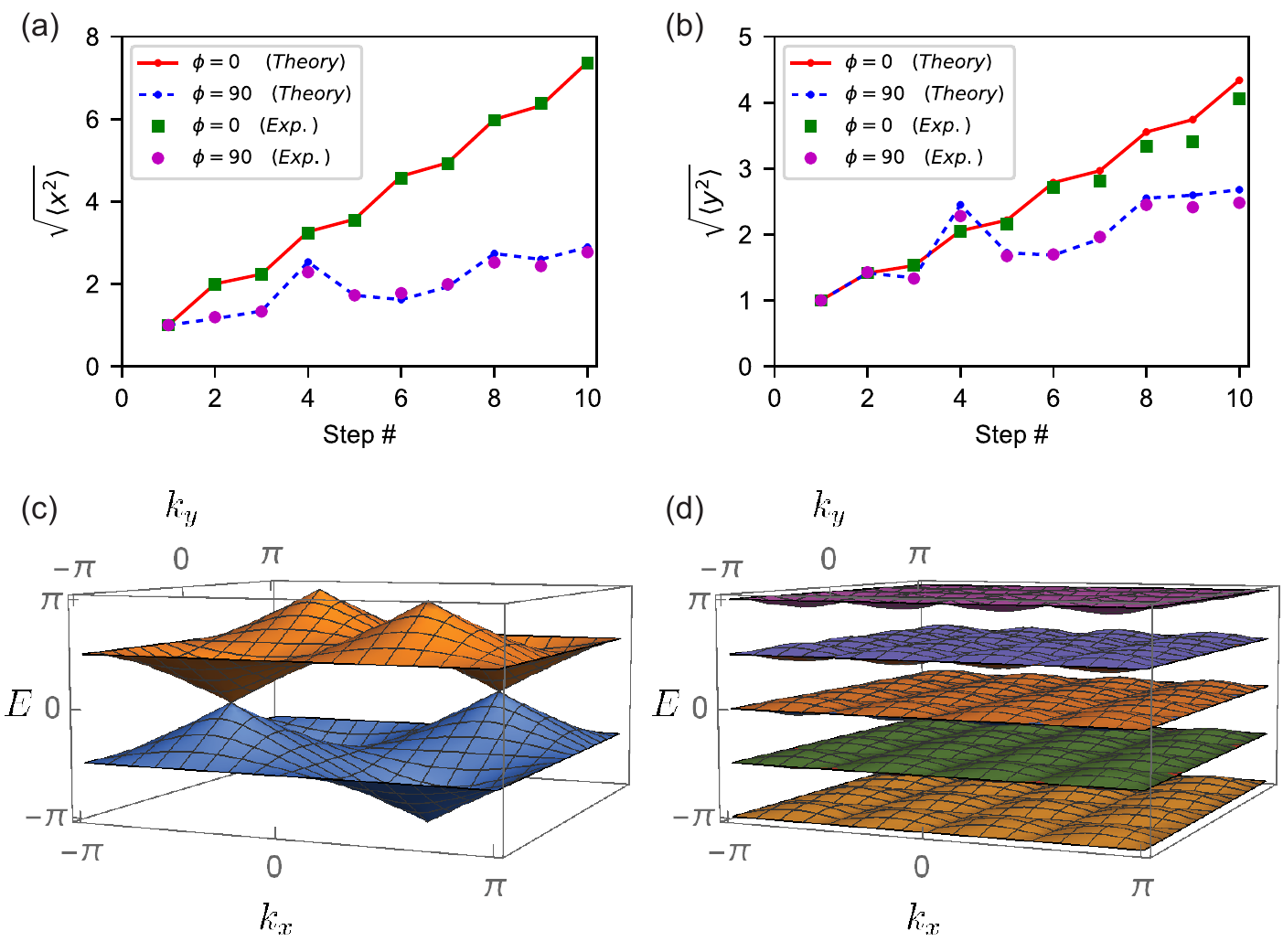}
\centering{}\caption{\label{Fig3}(a) Comparison of the theoretical and experimental results
for the variation of the random walk distribution quadratic mean in
the $x$ direction under no gauge field as well as under the gauge
field with $\phi=\frac{\pi}{2}$. (b) Comparison of the theoretical
and experimental results for the variation of the random walk distribution
quadratic mean in the $y$ direction under no gauge field as well
as under the gauge field with $\phi=\frac{\pi}{2}$. (c, d) Band structure
of the system under no phase modulation ($\phi=0$) and a linear phase
modulation ($\phi=\frac{\pi}{2}$). }
\end{figure}

\twocolumngrid

To provide a more quantitative analysis of the effect of the gauge
field on the particle confinement, we calculate the variation of the
spatial quadratic means of the random walk distribution as a function
of time step. Figures \ref{fig:Fig3a} and \ref{fig:Fig3b} plot the
quadratic means in the $x$ and $y$ directions with gauge fields
of $\phi=0$ and $\phi=\pi/2$. With no applied gauge field, the quadratic
means show nearly linear variation with the time step, consistent
with ballistic diffusion (See Supplementary Material \cite{supplem}
for analytical explanation). But under the application of the gauge
field with $\phi=\pi/2$, the quadratic means show reduced diffusion.
The decrease of the quadratic means in both directions is due to the
confinement of the random walk\textquoteright s distribution under
a constant pseudo magnetic field. Figures \ref{fig:Fig3a} and \ref{fig:Fig3b}
confirm the agreement of the experimental results with the theoretical
predictions, both with and without the effect of the gauge field.
(See also Fig. S3 for the numerical study of the variation of quadratic
means over a larger number of steps)

In order to better understand the confinement of the random walker
in the presence of a gauge field, we first calculate the band structure
of the random walk. The full evolution of the walker is determined
by the single-step propagation matrix $U$, which advances the random
walker distribution by one time-step. According to Floquet band theory,
the single-step propagation matrix determines the effective Hamiltonian
from $U=e^{-iH_{eff}}$, which gives the band structure of the walker.
With no synthetic gauge field ($\phi=0$), we can analytically solve
for the dispersion relation of the walker (See theoretical analysis
section in the Supplementary Material \cite{supplem}). The quasi-energy
bands of the system in this case are

\begin{equation}
E_{\pm}=\pm\arccos\left(\sin\left(k_{x}\right)\sin\left(k_{y}\right)\right),
\end{equation}
where $k_{x}$ and $k_{y}$ are the momentum wave vectors in inverse
synthetic space. Figure \ref{fig:Fig3c} shows the corresponding band
structure of the system. Because of the discrete nature of the random
walk, the quasi-energy spectrum wraps every $2\pi$, and therefore
we restrict the quasi-energies to the range of $-\pi$ and $\pi$.
As Fig. \ref{fig:Fig3c} shows, the system is gapless, and there are
four Dirac points, two at $E=0$ and two at $E=\pm\pi$. 

We next consider the effect of the synthetic gauge field on the band
structure. Figure \ref{fig:Fig3d} shows the band diagram for the
case of $\phi=\frac{\pi}{2}$. An analytical solution for this case
also exists (see Supplementary Material \cite{supplem}), with a quasi-energy
band structure given by 

\begin{equation}
E_{n,\pm}=\frac{n\pi}{2}\pm\frac{1}{4}\arccos\left(1-\frac{1}{2}\sin^{2}\left(2k_{x}\right)\sin^{2}\left(2k_{y}\right)\right)
\end{equation}
for $n\in\mathbb{Z}$. Similar to the case of integer quantum Hall
effect, the introduction of a gauge field produces a series of topological
bands. For $\phi=\frac{\pi}{2}$, we observe four doubly degenerate
bands. However, because of the wrapping of pseudo-energy, one set
of bands is split and appears close to energies $\pm\pi$. In contrast
to the zero gauge field, the band structure in the presence of a synthetic
gauge field exhibits bandgaps that lead to the confinement of the
random walker. We have also obtained the corresponding band diagrams
for several other choices of $\phi$. We have presented these results
(See Fig. S2) along with their derivation in the Supplementary Material
\cite{supplem}. 

\begin{figure}
\subfloat[\label{fig:Fig4a}]{}\subfloat[\label{fig:Fig4b}]{}\subfloat[\label{fig:Fig4c}]{}\subfloat[\label{fig:Fig4d}]{}\includegraphics{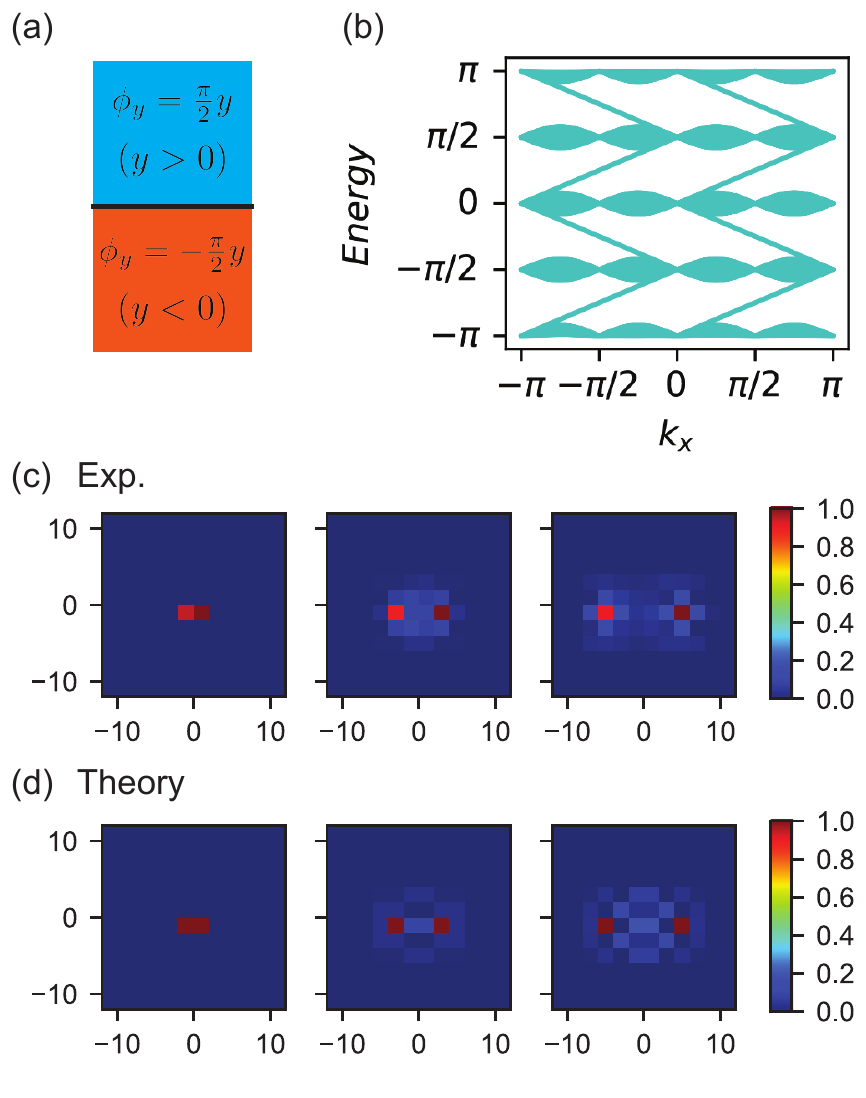}\caption{\label{Fig4}(a) The schematic describing the phase modulation pattern
of $\phi_{y}=\left|y\right|\phi$ for $\phi=\frac{\pi}{2}$ in the
synthetic space. (b) Band diagram of the corresponding system which
clearly shows the presence of edge states in the bandgap. (c) Experimental
observations and (d) theoretical predictions of the evolution of the
random walk distribution moving along the boundary under the phase
modulation of $\phi_{y}=\left|y\right|\phi$ for $\phi=\frac{\pi}{2}$.
The left, middle, and right columns show the distributions at time
steps of 1, 5, and 9, respectively. In these plots all the distributions
are normalized to their maximum. }
\end{figure}

One consequence of a gauge field is the presence of edge states at
the boundaries. In this synthetic space, we can make a boundary by
applying two different gauge fields to two neighboring regions. Here,
using a phase modulation pattern of $\phi_{y}=y\phi$ for $y>0$ and
$\phi_{y}=-y\phi$ for $y<0$, we realize two domains with opposite
magnetic fields ($y>0$ and $y<0$), as illustrated in Fig. \ref{fig:Fig4a}.
Figure \ref{fig:Fig4b} shows the band structure for such a phase
pattern with $\phi=\frac{\pi}{2}$. The band diagram contains multiple
bandgaps hosting unidirectional edge states that propagate at the
boundary in opposite directions. The corresponding band diagrams for
several other choices of $\phi$ are also presented in the Supplementary
Material \cite{supplem}(See Fig. S4). 

Figure \ref{fig:Fig4c} shows experimentally measured results for
the phase modulation pattern shown in Fig. \ref{fig:Fig4a}. We start
the random walker at the interface between the two magnetic domains,
precisely where edge states should be present. In this case the random
walker predominantly walks along the edge, remaining confined to the
boundary between the two regions. These results are consistent with
the numerical simulations demonstrating how the edge states cause
the random walk distribution to move mainly along the boundary (Fig.
\ref{fig:Fig4d}).

Typical topological quantum walks result in unidirectional edge state
propagation. Here, however, we do not see unidirectional movements
because we are initializing the walker at a position eigenstate, which
is a superposition of all energy eigenstates of the band structure.
As can be seen from Fig. \ref{fig:Fig4b}, different energy bands
support topological edge states propagating in either the left or
right direction. We could excite specific edge modes by engineering
the initial distribution of the random walk to be confined in corresponding
energies. Additionally, more complicated phase modulation patterns
can be harnessed to produce sharp edges in the synthetic space. 

In conclusion, we have implemented time-multiplexed two-dimensional
quantum random walks with a synthetic gauge field. This gauge field
leads to the confinement of the walker evolution. Through application
of an in-homogeneous gauge field on this random walk, we observed
the creation of topological edge states that are confined at the boundary
of two distinct gauge fields. These results demonstrate a versatile
approach to create various types of band structures with tunable number
of bandgaps. In order to increase the number of steps, we used optical
amplifiers to compensate for round-trip losses without damaging the
phase coherence of the optical pulses. These losses can be reduced
by decreasing coupler losses through fiber splicing and the use of
modulators with lower insertion loss. Eliminating these losses opens
up a path towards quantum random walks that can be implemented at
the single photon level, or in higher dimensions. Addition of optical
nonlinearities and integration of this platform with single photon
emitters could provide another interesting opportunity to study topological
band structures with optical interactions \cite{Chalabi2018,PhysRevLett.116.093601}.
Ultimately, our results expand the toolbox of quantum photonic simulation
and provide a scalable architecture to study photonic random walks
with non-trivial topologies.

This work was supported by the Air Force Office of Scientific Research-Multidisciplinary
University Research Initiative (Grant FA9550-16-1-0323), the Physics
Frontier Center at the Joint Quantum Institute, the National Science
Foundation (Grant No. PHYS. 1415458), and the Center for Distributed
Quantum Information. The authors would also like to acknowledge support
from the U.S. Department of Defense. 

\bibliographystyle{apsrev4-1}
\bibliography{refers_arXiv}

\onecolumngrid



\section*{Supplementary material}

# Supplementary Information: A synthetic gauge field for two-dimensional time-multiplexed quantum random walks

In this supplementary material, we first explain the experimental details in the first section. The details of the theoretical analysis are summarized in the subsequent section.

## I. EXPERIMENTAL DETAILS:

For the purpose of pulse generation in this random walk experiment, we have used a laser diode operating at the C-band of the telecom wavelength (1550nm). By modulation of this laser made by Bookham technology (LC25W5172BA-J34), we have generated pulses with the power of $10mW$ and the pulse duration of two nanoseconds. These pulses are sent into the setup with the repetition rate of $10\mu s$. We have used these pulses to initiate the random walk in our synthetic two dimensional space. Moreover, the polarization of the incident pulse is controlled using the polarization beam splitter.

Figure 1b in the manuscript shows the experimental setup used for making the synthetic two dimensional space. The initial pulse after passing through the first 50/50 beam splitter chooses to continue its path through either the 1m or 2m fiber length. This leads to different delays that can be regarded as choosing left or right direction in the $x$ movement. By reaching the next beam splitter, the pulse chooses either the 130m or 100m fiber pool. This determines whether the pulse has gone to the up or down direction in the $y$ movement. By this way, we encode the $x$ and $y$ movement of the pulses based on the delay in reaching the detectors. We have used 90/10 beam splitters to out-couple 10% of the light for the detection purposes.

In order to compensate the loss in the experimental setup, we have used semiconductor optical amplifiers made by Thorlabs (SOA1117S). These amplifiers, which are turned on for a fraction a repetition period at each cycle, provide the capability to amplify the pulses without ruining their phase coherence. However, in addition to amplifying the signal they increase the background noise due to spontaneous emission. Such noises should be filtered out using band-pass filter. For this purpose, we have used narrow band-pass filters ($< 0.3nm$) right after semiconductor optical amplifiers as shown in Fig. 1b.

Despite this compensation, the ratio of the optical pulse power at different positions of the synthetic space relative to the noise degrades with increasing time steps due to the diffusion and the added noise. Because of the degradation of the signal to noise ratio and the finite ratio of the time delays corresponding to $x$ and $y$ movements, we limited our measurement of the random walk distribution to 10 steps. By measuring the power of the pulses present in each step, we can determine the distribution of the random walk at every time step.

In addition, the polarization of the incident pulse varies by passing through the fiber network. To compensate these polarization changes, we have used polarization controllers in our setup.

Furthermore, we have used two phase modulators in this setup that apply opposite phases to the left and right moving pulses. These phase modulators (MPZ-LN-10-P-P-FA-FA) made by Cybel company can provide switching speed of $12Gb/s$ and have $3dB$ loss.

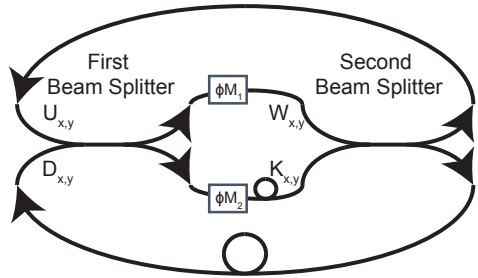

Figure S1. Simplified version of the schematic of the 2D random walk setup.

## II.  THEORETICAL ANALYSIS:

In the followings, the evolution of pulses (representing the random walkers) in the two-dimensional (2D) random walk is analyzed. For the theoretical analysis of the random walk, we can assume that all the losses in the setup including the losses caused by 90/10 beam splitters as well as phase modulators are fully compensated by the amplifiers. Moreover, we can assume that polarization controllers compensate all the polarization changes in the setup. Therefore, for theoretical analysis we can simplify the schematic and consider an ideal setup as shown in Fig. S1.

Two pulses just before the first beam splitter, $\begin{bmatrix} U_{x,y}^{(n)} \\ D_{x,y}^{(n)} \end{bmatrix}$, will produce the output pulses as (movement in $x$ direction):

$$\begin{bmatrix} W_{x-1,y} \\ K_{x+1,y} \end{bmatrix} = \frac{1}{\sqrt{2}} \begin{bmatrix} e^{i\phi_{n,y}} & 0 \\ 0 & e^{-i\phi_{n,y}} \end{bmatrix} \begin{bmatrix} 1 & -1 \\ 1 & 1 \end{bmatrix} \begin{bmatrix} U_{x,y}^{(n)} \\ D_{x,y}^{(n)} \end{bmatrix}.$$

Note that the applied phases are assumed to be explicit functions of time step ($n$) as well as $y$ coordinate but not $x$ coordinate due to experimental limitations.

Pulses which are input to the second beam splitter, $\begin{bmatrix} W_{x,y} \\ K_{x,y} \end{bmatrix}$, will produce the outputs as (movement in $y$ direction):

$$\begin{bmatrix} U_{x,y+1}^{(n+1)} \\ D_{x,y-1}^{(n+1)} \end{bmatrix} = \frac{1}{\sqrt{2}} \begin{bmatrix} 1 & -1 \\ 1 & 1 \end{bmatrix} \begin{bmatrix} W_{x,y} \\ K_{x,y} \end{bmatrix}$$

By considering the pulse of $\begin{bmatrix} W_{x-1,y} \\ 0 \end{bmatrix}$ caused by $\begin{bmatrix} U_{x,y}^{(n)} \\ D_{x,y}^{(n)} \end{bmatrix}$, we have:

$$\begin{aligned} \begin{bmatrix} U_{x-1,y+1}^{(n+1)} \\ D_{x-1,y-1}^{(n+1)} \end{bmatrix} &= \frac{1}{\sqrt{2}} \begin{bmatrix} 1 & -1 \\ 1 & 1 \end{bmatrix} \begin{bmatrix} W_{x-1,y} \\ 0 \end{bmatrix} \\ &= \frac{1}{2} \begin{bmatrix} 1 & -1 \\ 1 & 1 \end{bmatrix} \begin{bmatrix} 1 & 0 \\ 0 & 0 \end{bmatrix} \begin{bmatrix} e^{i\phi_{n,y}} & 0 \\ 0 & e^{-i\phi_{n,y}} \end{bmatrix} \begin{bmatrix} 1 & -1 \\ 1 & 1 \end{bmatrix} \begin{bmatrix} U_{x,y}^{(n)} \\ D_{x,y}^{(n)} \end{bmatrix} \\ &= \frac{1}{2} \begin{bmatrix} e^{i\phi_{n,y}} & -e^{i\phi_{n,y}} \\ e^{i\phi_{n,y}} & -e^{i\phi_{n,y}} \end{bmatrix} \begin{bmatrix} U_{x,y}^{(n)} \\ D_{x,y}^{(n)} \end{bmatrix} \end{aligned}$$

Furthermore, by considering the pulse of $\begin{bmatrix} 0 \\ K_{x+1,y} \end{bmatrix}$ caused by $\begin{bmatrix} U_{x,y}^{(n)} \\ D_{x,y}^{(n)} \end{bmatrix}$, we have:

$$\begin{aligned} \begin{bmatrix} U_{x+1,y+1}^{(n+1)} \\ D_{x+1,y-1}^{(n+1)} \end{bmatrix} &= \frac{1}{\sqrt{2}} \begin{bmatrix} 1 & -1 \\ 1 & 1 \end{bmatrix} \begin{bmatrix} 0 \\ K_{x+1,y} \end{bmatrix} \\ &= \frac{1}{2} \begin{bmatrix} 1 & -1 \\ 1 & 1 \end{bmatrix} \begin{bmatrix} 0 & 0 \\ 0 & 1 \end{bmatrix} \begin{bmatrix} e^{i\phi_{n,y}} & 0 \\ 0 & e^{-i\phi_{n,y}} \end{bmatrix} \begin{bmatrix} 1 & -1 \\ 1 & 1 \end{bmatrix} \begin{bmatrix} U_{x,y}^{(n)} \\ D_{x,y}^{(n)} \end{bmatrix} \\ &= \frac{1}{2} \begin{bmatrix} -e^{-i\phi_{n,y}} & -e^{-i\phi_{n,y}} \\ e^{-i\phi_{n,y}} & e^{-i\phi_{n,y}} \end{bmatrix} \begin{bmatrix} U_{x,y}^{(n)} \\ D_{x,y}^{(n)} \end{bmatrix} \end{aligned}$$

Therefore, $\begin{bmatrix} U_{x,y}^{(n)} \\ D_{x,y}^{(n)} \end{bmatrix}$ will produce the following pulses after traversing both of the beam splitters:

$$U_{x-1,y+1}^{(n+1)} = \frac{1}{2} \left( e^{i\phi_{n,y}} U_{x,y}^{(n)} - e^{i\phi_{n,y}} D_{x,y}^{(n)} \right)$$

$$U_{x+1,y+1}^{(n+1)} = \frac{1}{2} \left( -e^{-i\phi_{n,y}} U_{x,y}^{(n)} - e^{-i\phi_{n,y}} D_{x,y}^{(n)} \right)$$

$$D_{x-1,y-1}^{(n+1)} = \frac{1}{2}\left(e^{i\phi_{n,y}}U_{x,y}^{(n)} - e^{i\phi_{n,y}}D_{x,y}^{(n)}\right)$$

$$D_{x+1,y-1}^{(n+1)} = \frac{1}{2}\left(e^{-i\phi_{n,y}}U_{x,y}^{(n)} + e^{-i\phi_{n,y}}D_{x,y}^{(n)}\right)$$

Alternatively, based on the above results $\begin{bmatrix} U_{x,y}^{(n+1)} \\ D_{x,y}^{(n+1)} \end{bmatrix}$ can be produced from other pulses as:

$$U_{x,y}^{(n+1)} = \frac{e^{i\phi_{n,y-1}}}{2}\left(U_{x+1,y-1}^{(n)} - D_{x+1,y-1}^{(n)}\right) - \frac{e^{-i\phi_{n,y-1}}}{2}\left(U_{x-1,y-1}^{(n)} + D_{x-1,y-1}^{(n)}\right)$$

$$D_{x,y}^{(n+1)} = \frac{e^{i\phi_{n,y+1}}}{2}\left(U_{x+1,y+1}^{(n)} - D_{x+1,y+1}^{(n)}\right) + \frac{e^{-i\phi_{n,y+1}}}{2}\left(U_{x-1,y+1}^{(n)} + D_{x-1,y+1}^{(n)}\right)$$

The above two equations fully describe the evolution of pulses by time. We can prove that any constant phase independent of time and coordinate will not change the dynamics of evolution. For this purpose, we separate out any such constant phase from $\phi_{n,y}$ and we assume that $\phi_{n,y}$ can be written as:

$$\phi_{n,y} = \Phi + \phi'_{n,y}$$

Therefore:

$$U_{x,y}^{(n+1)} = \frac{e^{i\Phi+i\phi'_{n,y-1}}}{2}\left(U_{x+1,y-1}^{(n)} - D_{x+1,y-1}^{(n)}\right) - \frac{e^{-i\Phi-i\phi'_{n,y-1}}}{2}\left(U_{x-1,y-1}^{(n)} + D_{x-1,y-1}^{(n)}\right)$$

$$D_{x,y}^{(n+1)} = \frac{e^{i\Phi+i\phi'_{n,y+1}}}{2}\left(U_{x+1,y+1}^{(n)} - D_{x+1,y+1}^{(n)}\right) + \frac{e^{-i\Phi-i\phi'_{n,y+1}}}{2}\left(U_{x-1,y+1}^{(n)} + D_{x-1,y+1}^{(n)}\right)$$

By defining $U'$ and $D'$ from $U$ and $D$ as:

$$U_{x,y}'^{(n)} = e^{ix\Phi}U_{x,y}^{(n)}$$

$$D_{x,y}'^{(n)} = e^{ix\Phi}D_{x,y}^{(n)}$$

We have:

$$U_{x,y}'^{(n+1)} = \frac{e^{i\phi'_{n,y-1}}}{2}\left(U_{x+1,y-1}'^{(n)} - D_{x+1,y-1}'^{(n)}\right) - \frac{e^{-i\phi'_{n,y-1}}}{2}\left(U_{x-1,y-1}'^{(n)} + D_{x-1,y-1}'^{(n)}\right)$$

$$D_{x,y}'^{(n+1)} = \frac{e^{i\phi'_{n,y+1}}}{2}\left(U_{x+1,y+1}'^{(n)} - D_{x+1,y+1}'^{(n)}\right) + \frac{e^{-i\phi'_{n,y+1}}}{2}\left(U_{x-1,y+1}'^{(n)} + D_{x-1,y+1}'^{(n)}\right)$$

This shows that the evolution is invariant under such a constant phase application.

By defining $S_{x,y}^{(n)} = U_{x,y}^{(n)} + D_{x,y}^{(n)}$ and $P_{x,y}^{(n)} = U_{x,y}^{(n)} - D_{x,y}^{(n)}$, the obtained equations can be written as:

$$\begin{bmatrix} S_{x,y}^{(n+1)} \\ P_{x,y}^{(n+1)} \end{bmatrix} = \frac{1}{2}\begin{bmatrix} e^{-i\phi_{n,y+1}} & e^{i\phi_{n,y+1}} & -e^{-i\phi_{n,y-1}} & e^{i\phi_{n,y-1}} \\ -e^{-i\phi_{n,y+1}} & -e^{i\phi_{n,y+1}} & -e^{-i\phi_{n,y-1}} & e^{i\phi_{n,y-1}} \end{bmatrix}\begin{bmatrix} S_{x-1,y+1}^{(n)} \\ P_{x+1,y+1}^{(n)} \\ S_{x-1,y-1}^{(n)} \\ P_{x+1,y-1}^{(n)} \end{bmatrix}$$

Now that we have the governing equation in the real space, we can think about its solution in the Fourier space. By defining $s^{(n)}_{k_x,k_y}$ and $p^{(n)}_{k_x,k_y}$ as Fourier transforms of $S^{(n)}_{x,y}$ and $P^{(n)}_{x,y}$, we have:

$$\begin{bmatrix} S^{(n)}_{x,y} \\ P^{(n)}_{x,y} \end{bmatrix} = \frac{1}{4\pi^2} \begin{bmatrix} \int \int_{k_x,k_y} s^{(n)}_{k_x,k_y} e^{ik_x x + ik_y y} dk_x dk_y \\ \int \int_{k_x,k_y} p^{(n)}_{k_x,k_y} e^{ik_x x + ik_y y} dk_x dk_y \end{bmatrix}$$

Therefore:

$$\begin{bmatrix} \int \int_{k_x,k_y} s^{(n+1)}_{k_x,k_y} e^{ik_x x + ik_y y} dk_x dk_y \\ \int \int_{k_x,k_y} p^{(n+1)}_{k_x,k_y} e^{ik_x x + ik_y y} dk_x dk_y \end{bmatrix} = \frac{1}{2} \begin{bmatrix} e^{-i\phi_{n,y+1}} & e^{i\phi_{n,y+1}} & -e^{-i\phi_{n,y-1}} & e^{i\phi_{n,y-1}} \\ -e^{-i\phi_{n,y+1}} & -e^{i\phi_{n,y+1}} & -e^{-i\phi_{n,y-1}} & e^{i\phi_{n,y-1}} \end{bmatrix}$$

$$\times \begin{bmatrix} \int \int_{k_x,k_y} e^{ik_y - ik_x} s^{(n)}_{k_x,k_y} e^{ik_x x + ik_y y} dk_x dk_y \\ \int \int_{k_x,k_y} e^{ik_x + ik_y} p^{(n)}_{k_x,k_y} e^{ik_x x + ik_y y} dk_x dk_y \\ \int \int_{k_x,k_y} e^{-ik_x - ik_y} s^{(n)}_{k_x,k_y} e^{ik_x x + ik_y y} dk_x dk_y \\ \int \int_{k_x,k_y} e^{ik_x - ik_y} p^{(n)}_{k_x,k_y} e^{ik_x x + ik_y y} dk_x dk_y \end{bmatrix}$$

This equation can be used to solve the Fourier transforms as functions of the time step. In the following subsections, we solve this equation for two cases of no phase modulation as well as time independent linear phase modulation.

## A.  Zero phase modulation:

For the case of no applied phase, we have:

$$\begin{bmatrix} s^{(n+1)}_{k_x,k_y} \\ p^{(n+1)}_{k_x,k_y} \end{bmatrix} = \begin{bmatrix} ie^{-ik_x}\sin(k_y) & e^{ik_x}\cos(k_y) \\ -e^{-ik_x}\cos(k_y) & -ie^{ik_x}\sin(k_y) \end{bmatrix} \begin{bmatrix} s^{(n)}_{k_x,k_y} \\ p^{(n)}_{k_x,k_y} \end{bmatrix}$$

Note that the evolution matrix has the determinant of 1. Based on this matrix, the effective Hamiltonian, $H_{eff} = i \log(U)$, is given by:

$$H = \begin{pmatrix} -\cos(k_x)\sin(k_y) & ie^{ik_x}\cos(k_y) \\ -ie^{-ik_x}\cos(k_y) & \cos(k_x)\sin(k_y) \end{pmatrix} \frac{\arccos(\sin(k_x)\sin(k_y))}{\sin(\arccos(\sin(k_x)\sin(k_y)))}$$

The obtained Hamiltonian is obviously hermitian. The eigenvalues of this Hamiltonian are given by:

$$E_\pm = \pm \arccos(\sin(k_x)\sin(k_y))$$

Moreover, the eigenvectors are given by:

$$\begin{pmatrix} \cos(k_x)\sin(k_y) \mp \sqrt{1 - \sin^2(k_x)\sin^2(k_y)} \\ ie^{-ik_x}\cos(k_y) \end{pmatrix}$$

Therefore, by defining $S$ as

$$S = \begin{pmatrix} \cos(k_x)\sin(k_y) - \sqrt{1 - \sin^2(k_x)\sin^2(k_y)} & \cos(k_x)\sin(k_y) + \sqrt{1 - \sin^2(k_x)\sin^2(k_y)} \\ ie^{-ik_x}\cos(k_y) & ie^{-ik_x}\cos(k_y) \end{pmatrix},$$

we can write the Hamiltonian as:

$$H = S \begin{pmatrix} E_+ & 0 \\ 0 & E_- \end{pmatrix} S^{-1}$$

Based on the obtained eigen-energies, the band diagram for the zero phase modulation has been plotted in Fig. 3a. The evolution after $n$ steps is given by:

$$\begin{bmatrix} s^{(n)}_{k_x,k_y} \\ p^{(n)}_{k_x,k_y} \end{bmatrix} = \begin{bmatrix} ie^{-ik_x}\sin(k_y) & e^{ik_x}\cos(k_y) \\ -e^{-ik_x}\cos(k_y) & -ie^{ik_x}\sin(k_y) \end{bmatrix}^n \begin{bmatrix} s^{(0)}_{k_x,k_y} \\ p^{(0)}_{k_x,k_y} \end{bmatrix}$$

$$= U^n \begin{bmatrix} s^{(0)}_{k_x,k_y} \\ p^{(0)}_{k_x,k_y} \end{bmatrix}$$

The deterministic equation for the evolution matrix, $\lambda^2 - 2\lambda \sin(k_y)\sin(k_x) + 1 = 0$, gives the following Eigenvalues:

$$\lambda_{1,2} = \sin(k_y)\sin(k_x) \mp \sqrt{\sin^2(k_y)\sin^2(k_x) - 1}$$

They are equal to $e^{-iE_\pm}$ as we expect. We can define the following two orthogonal functions for performing the matrix power:

$$e^{(n)}_{k_x,k_y}, f^{(n)}_{k_x,k_y} = e^{-ik_x}\cos(k_y)\, s^{(n)}_{k_x,k_y} + \left(i\cos(k_x)\sin(k_y) \pm \sqrt{\sin^2(k_y)\sin^2(k_x) - 1}\right) p^{(n)}_{k_x,k_y}$$

These functions obey the following independent equations:

$$e_{k_x,k_y}^{(n+1)} = \left(\sin\left(k_x\right)\sin\left(k_y\right) - \sqrt{\sin^2\left(k_y\right)\sin^2\left(k_x\right) - 1}\right) e_{k_x,k_y}^{(n)}$$

$$f_{k_x,k_y}^{(n+1)} = \left(\sin\left(k_x\right)\sin\left(k_y\right) + \sqrt{\sin^2\left(k_y\right)\sin^2\left(k_x\right) - 1}\right) f_{k_x,k_y}^{(n)}$$

Assuming that the whole evolution is caused by a single pulse at the origin ($U_{x,y}^{(0)} = \delta\left(x\right)\delta\left(y\right)$, $D_{x,y}^{(0)} = 0$ which is equivalent to $S_{x,y}^{(0)} = \delta\left(x\right)\delta\left(y\right)$, $P_{x,y}^{(0)} = \delta\left(x\right)\delta\left(y\right)$), then we would have (the Fourier transform of Dirac delta function is constant unity):

$$\begin{bmatrix} s_{k_x,k_y}^{(0)} \\ p_{k_x,k_y}^{(0)} \end{bmatrix} = \begin{bmatrix} 1 \\ 1 \end{bmatrix}$$

Therefore, for the down channel pulses, $d_{k_x,k_y}^{(n)} = 0.5\left(s_{k_x,k_y}^{(n)} - p_{k_x,k_y}^{(n)}\right)$, we have:

$$d_{k_x,k_y}^{(n)} = 0.5\left(s_{k_x,k_y}^{(n)} - p_{k_x,k_y}^{(n)}\right)$$
$$= e^{ik_y}\cos\left(k_x\right)\frac{\sin\left(n\arccos\left(\sin\left(k_x\right)\sin\left(k_y\right)\right)\right)}{\sin\left(\arccos\left(\sin\left(k_x\right)\sin\left(k_y\right)\right)\right)}$$

All the moments such as distribution variances and averages ($\left\langle x^2\right\rangle_D$, $\left\langle y^2\right\rangle_D$, $\left\langle x\right\rangle_D$ and $\left\langle y\right\rangle_D$) as functions of time step $n$, can be obtained from the above expression.

Based on the fact that:

$$d_{k_x,k_y}^{(n)} = \sum_{x,y} D_{x,y}^{(n)} e^{-ik_x x} e^{-ik_y y}$$

Then:

$$d_{l_x,l_y}^{(n)*} d_{k_x,k_y}^{(n)} = \sum_{p,q}\sum_{x,y} e^{-i(k_x x - l_x p)} e^{-i(k_y y - l_y q)} D_{p,q}^{(n)*} D_{x,y}^{(n)}$$

Therefore, the following equations can be obtained, straightforwardly:

$$P_D = \sum_{x,y}\left|D_{x,y}^{(n)}\right|^2 = \frac{1}{4\pi^2}\int_{k_y=0}^{2\pi}\int_{k_x=0}^{2\pi}\left|d_{k_x,k_y}^{(n)}\right|^2 dk_x dk_y$$
$$= \frac{1}{4\pi^2}\int_{k_y=0}^{2\pi}\int_{k_x=0}^{2\pi}\cos^2\left(k_x\right)\frac{\sin^2\left(n\arccos\left(\sin\left(k_x\right)\sin\left(k_y\right)\right)\right)}{\sin^2\left(\arccos\left(\sin\left(k_x\right)\sin\left(k_y\right)\right)\right)} dk_x dk_y$$

$$\left\langle x\right\rangle_D = \sum_{x,y} x\left|D_{x,y}^{(n)}\right|^2 = \frac{1}{4\pi^2}\int_{k_y=0}^{2\pi}\int_{k_x=0}^{2\pi} d_{k_x,k_y}^{(n)*}\left(i\frac{d}{dk_x}\right) d_{k_x,k_y}^{(n)} dk_x dk_y = 0$$

$$\left\langle y\right\rangle_D = \sum_{x,y} y\left|D_{x,y}^{(n)}\right|^2 = \frac{1}{4\pi^2}\int_{k_y=0}^{2\pi}\int_{k_x=0}^{2\pi} d_{k_x,k_y}^{(n)*}\left(i\frac{d}{dk_y}\right) d_{k_x,k_y}^{(n)} dk_x dk_y$$
$$= \frac{-1}{4\pi^2}\int_{k_y=0}^{2\pi}\int_{k_x=0}^{2\pi}\cos^2\left(k_x\right)\frac{\sin^2\left(n\arccos\left(\sin\left(k_x\right)\sin\left(k_y\right)\right)\right)}{\sin^2\left(\arccos\left(\sin\left(k_x\right)\sin\left(k_y\right)\right)\right)} dk_x dk_y = -P_D$$

$$\langle x^2 \rangle_D = \sum_{x,y} x^2 \left| D_{x,y}^{(n)} \right|^2 = \frac{1}{4\pi^2} \int_{k_y=0}^{2\pi} \int_{k_x=0}^{2\pi} d_{k_x,k_y}^{(n)*} \left( i \frac{d}{dk_x} \right)^2 d_{k_x,k_y}^{(n)} dk_x dk_y$$

$$= \frac{n^2}{4\pi^2} \int_{k_y=0}^{2\pi} \int_{k_x=0}^{2\pi} \frac{\cos^4(k_x) \sin^2(k_y)}{\left(1 - \sin^2(k_x) \sin^2(k_y)\right)^2} \sin^2\left(n \cos^{-1}\left(\sin(k_x) \sin(k_y)\right)\right) dk_x dk_y$$

$$- \frac{3n}{8\pi^2} \int_{k_y=0}^{2\pi} \int_{k_x=0}^{2\pi} \frac{\sin(k_x) \sin(k_y) \cos^2(k_x) \cos^2(k_y)}{\left(1 - \sin^2(k_x) \sin^2(k_y)\right)^{5/2}} \sin\left(2n \cos^{-1}\left(\sin(k_x) \sin(k_y)\right)\right) dk_x dk_y$$

$$+ \frac{3}{4\pi^2} \int_{k_y=0}^{2\pi} \int_{k_x=0}^{2\pi} \frac{\sin^2(k_x) \sin^2(k_y) \cos^2(k_x) \cos^2(k_y)}{\left(1 - \sin^2(k_x) \sin^2(k_y)\right)^3} \sin^2\left(n \cos^{-1}\left(\sin(k_x) \sin(k_y)\right)\right) dk_x dk_y$$

$$+ \frac{1}{4\pi^2} \int_{k_y=0}^{2\pi} \int_{k_x=0}^{2\pi} \frac{\cos^2(k_x) \cos^2(k_y)}{\left(1 - \sin^2(k_x) \sin^2(k_y)\right)^2} \sin^2\left(n \cos^{-1}\left(\sin(k_x) \sin(k_y)\right)\right) dk_x dk_y$$

$$\langle y^2 \rangle_D = \sum_{x,y} y^2 \left| D_{x,y}^{(n)} \right|^2 = \frac{1}{4\pi^2} \int_{k_y=0}^{2\pi} \int_{k_x=0}^{2\pi} d_{k_x,k_y}^{(n)*} \left( i \frac{d}{dk_y} \right)^2 d_{k_x,k_y}^{(n)} dk_x dk_y$$

$$= \frac{n^2}{4\pi^2} \int_{k_y=0}^{2\pi} \int_{k_x=0}^{2\pi} \frac{\sin^2(k_x) \cos^2(k_x) \cos^2(k_y)}{\left(1 - \sin^2(k_x) \sin^2(k_y)\right)^2} \sin^2\left(n \cos^{-1}\left(\sin(k_x) \sin(k_y)\right)\right) dk_x dk_y$$

$$+ \frac{n}{4\pi^2} \int_{k_y=0}^{2\pi} \int_{k_x=0}^{2\pi} \frac{\cos^2(k_x) \sin(k_x) \sin(k_y)}{\left(1 - \sin^2(k_x) \sin^2(k_y)\right)^{3/2}} \sin\left(2n \cos^{-1}\left(\sin(k_x) \sin(k_y)\right)\right) dk_x dk_y$$

$$- \frac{3n}{8\pi^2} \int_{k_y=0}^{2\pi} \int_{k_x=0}^{2\pi} \frac{\cos^4(k_x) \sin(k_x) \sin(k_y)}{\left(1 - \sin^2(k_x) \sin^2(k_y)\right)^{5/2}} \sin\left(2n \cos^{-1}\left(\sin(k_x) \sin(k_y)\right)\right) dk_x dk_y$$

$$+ \frac{1}{2\pi^2} \int_{k_y=0}^{2\pi} \int_{k_x=0}^{2\pi} \frac{\cos^2(k_x) \left(\sin^2(k_x) \cos^2(k_y) - 1\right)}{\left(1 - \sin^2(k_x) \sin^2(k_y)\right)^2} \sin^2\left(n \cos^{-1}\left(\sin(k_x) \sin(k_y)\right)\right) dk_x dk_y$$

$$+ \frac{3}{4\pi^2} \int_{k_y=0}^{2\pi} \int_{k_x=0}^{2\pi} \frac{\cos^4(k_x)}{\left(1 - \sin^2(k_x) \sin^2(k_y)\right)^3} \sin^2\left(n \cos^{-1}\left(\sin(k_x) \sin(k_y)\right)\right) dk_x dk_y$$

In the limit of large $n$, we have:

$$P_D \to \frac{1}{8\pi^2} \int_{k_y=0}^{2\pi} \int_{k_x=0}^{2\pi} \frac{\cos^2(k_x)}{\sin^2\left(\arccos\left(\sin(k_x) \sin(k_y)\right)\right)} dk_x dk_y = \frac{1}{\pi}$$

$$\langle x \rangle_D \to 0$$

$$\langle y \rangle_D \to -\frac{1}{\pi}$$

$$\langle x^2 \rangle_D \to \frac{n^2}{8\pi^2} \int_{k_y=0}^{2\pi} \int_{k_x=0}^{2\pi} \frac{\cos^4(k_x) \sin^2(k_y)}{\left(1 - \sin^2(k_x) \sin^2(k_y)\right)^2} dk_x dk_y = \frac{n^2}{2\pi}$$

$$\langle y^2 \rangle_D \to \frac{n^2}{8\pi^2} \int_{k_y=0}^{2\pi} \int_{k_x=0}^{2\pi} \frac{\sin^2(k_x) \cos^2(k_x) \cos^2(k_y)}{\left(1 - \sin^2(k_x) \sin^2(k_y)\right)^2} dk_x dk_y = \frac{n^2}{6\pi}$$

which shows that $\langle y^2 \rangle_D$ is three times less than $\langle x^2 \rangle_D$. These results prove that the spatial quadratic means of the random walk distribution linearly vary with the time step. Note that these average values are normalized relative to

the total power in the up and down channels. However, they can also be normalized relative to the total power present in the corresponding channel. Such a normalization has been used in the manuscript to calculate the experimental and theoretical quadratic means. Since the probabilities of $P_D$ and $P_U$ tend toward constant values, by the latter normalization the asymptotic behavior of the quadratic means remains linear relative to the time step.

It is interesting to test some first steps:

$$d_{k_x,k_y}^{(0)} = 0 \rightarrow D_{x,y}^{(0)} = 0$$

$$d_{k_x,k_y}^{(1)} = e^{ik_y}\cos(k_x) \rightarrow D_{x,y}^{(1)} = \begin{cases} 0.5 \, (x=-1, y=-1) \\ 0.5 \, (x=1, y=-1) \\ 0 \, (Otherwise) \end{cases}$$

$$d_{k_x,k_y}^{(2)} = e^{ik_y}\sin(2k_x)\sin(k_y) \rightarrow D_{x,y}^{(2)} = \begin{cases} -0.25 \, (x=-2, y=-2) \\ 0 \, (x=0, y=-2) \\ 0.25 \, (x=2, y=-2) \\ 0.25 \, (x=-2, y=0) \\ 0 \, (x=0, y=0) \\ -0.25 \, (x=2, y=0) \\ 0 \, (Otherwise) \end{cases}$$

$$d_{k_x,k_y}^{(3)} = e^{ik_y}\cos(k_x)\left(\cos(2k_x)\cos(2k_y)-\cos(2k_x)-\cos(2k_y)\right) \rightarrow D_{x,y}^{(3)} = \begin{cases} 0.125 \, (x=-3, y=-3) \\ -0.125 \, (x=-1, y=-3) \\ -0.125 \, (x=1, y=-3) \\ 0.125 \, (x=3, y=-3) \\ -0.25 \, (x=-3, y=-1) \\ -0.25 \, (x=-1, y=-1) \\ -0.25 \, (x=1, y=-1) \\ -0.25 \, (x=3, y=-1) \\ 0.125 \, (x=-3, y=1) \\ -0.125 \, (x=-1, y=1) \\ -0.125 \, (x=1, y=1) \\ 0.125 \, (x=3, y=1) \\ 0 \, (Otherwise) \end{cases}$$

Similarly, we can investigate the up channel:

$$u_{k_x,k_y}^{(n)} = 0.5\left(s_{k_x,k_y}^{(n)} + p_{k_x,k_y}^{(n)}\right)$$

$$= \cos\left(n\arccos\left(\sin(k_x)\sin(k_y)\right)\right) + i\cos(k_y)\sin(k_x)\frac{\sin\left(n\arccos\left(\sin(k_x)\sin(k_y)\right)\right)}{\sin\left(\arccos\left(\sin(k_x)\sin(k_y)\right)\right)}$$

$$P_U = \sum_{x,y}\left|U_{x,y}^{(n)}\right|^2 = \frac{1}{4\pi^2}\int_{k_y=0}^{2\pi}\int_{k_x=0}^{2\pi}\left|u_{k_x,k_y}^{(n)}\right|^2 dk_x dk_y$$

$$= \frac{1}{4\pi^2}\int_{k_y=0}^{2\pi}\int_{k_x=0}^{2\pi}\cos^2\left(n\arccos\left(\sin(k_x)\sin(k_y)\right)\right) dk_x dk_y$$

$$+ \frac{1}{4\pi^2}\int_{k_y=0}^{2\pi}\int_{k_x=0}^{2\pi}\cos^2(k_y)\sin^2(k_x)\frac{\sin^2\left(n\arccos\left(\sin(k_x)\sin(k_y)\right)\right)}{\sin^2\left(\arccos\left(\sin(k_x)\sin(k_y)\right)\right)} dk_x dk_y$$

$$\langle x\rangle_U = \sum_{x,y} x\left|U_{x,y}^{(n)}\right|^2 = \frac{1}{4\pi^2}\int_{k_y=0}^{2\pi}\int_{k_x=0}^{2\pi}u_{k_x,k_y}^{(n)*}\left(i\frac{d}{dk_x}\right)u_{k_x,k_y}^{(n)} dk_x dk_y = 0$$

$$\langle y \rangle_U = \sum_{x,y} y \left| U_{x,y}^{(n)} \right|^2 = \frac{1}{4\pi^2} \int_{k_y=0}^{2\pi} \int_{k_x=0}^{2\pi} u_{k_x,k_y}^{(n)*} \left( i \frac{d}{dk_y} \right) u_{k_x,k_y}^{(n)} dk_x dk_y$$

$$= \left( 1 - \frac{2}{\pi} \right) n + \frac{1}{8\pi^2} \int_{k_y=0}^{2\pi} \int_{k_x=0}^{2\pi} \frac{\cos^2(k_x) \sin(k_x) \sin(k_y)}{\left( 1 - \sin^2(k_x) \sin^2(k_y) \right)^{3/2}} \sin\left( 2n \arccos\left( \sin(k_x) \sin(k_y) \right) \right) dk_x dk_y$$

$$\langle x^2 \rangle_U = \sum_{x,y} x^2 \left| U_{x,y}^{(n)} \right|^2 = \frac{1}{4\pi^2} \int_{k_y=0}^{2\pi} \int_{k_x=0}^{2\pi} u_{k_x,k_y}^{(n)*} \left( i \frac{d}{dk_x} \right)^2 u_{k_x,k_y}^{(n)} dk_x dk_y$$

$$= \frac{n^2}{8\pi^2} \int_{k_y=0}^{2\pi} \int_{k_x=0}^{2\pi} \frac{\cos^2(k_x) \sin^2(k_y) \left( 1 + \sin^2(k_x) \cos(2k_y) \right)}{\left( 1 - \sin^2(k_x) \sin^2(k_y) \right)^2} dk_x dk_y$$

$$+ \frac{n^2}{8\pi^2} \int_{k_y=0}^{2\pi} \int_{k_x=0}^{2\pi} \frac{\cos^4(k_x) \sin^2(k_y)}{\left( 1 - \sin^2(k_x) \sin^2(k_y) \right)^2} \cos\left( 2n \arccos\left( \sin(k_x) \sin(k_y) \right) \right) dk_x dk_y$$

$$+ \frac{3n}{8\pi^2} \int_{k_y=0}^{2\pi} \int_{k_x=0}^{2\pi} \frac{\sin(k_x) \sin(k_y) \cos^2(k_x) \cos^2(k_y)}{\left( 1 - \sin^2(k_x) \sin^2(k_y) \right)^{5/2}} \sin\left( 2n \arccos\left( \sin(k_x) \sin(k_y) \right) \right) dk_x dk_y$$

$$+ \frac{1}{4\pi^2} \int_{k_y=0}^{2\pi} \int_{k_x=0}^{2\pi} \frac{\sin^2(k_x) \cos^4(k_y)}{\left( 1 - \sin^2(k_x) \sin^2(k_y) \right)^3} \sin^2\left( n \arccos\left( \sin(k_x) \sin(k_y) \right) \right) dk_x dk_y$$

$$- \frac{1}{2\pi^2} \int_{k_y=0}^{2\pi} \int_{k_x=0}^{2\pi} \frac{\sin^2(k_x) \sin^2(k_y) \cos^2(k_x) \cos^2(k_y)}{\left( 1 - \sin^2(k_x) \sin^2(k_y) \right)^3} \sin^2\left( n \arccos\left( \sin(k_x) \sin(k_y) \right) \right) dk_x dk_y$$

$$\langle y^2 \rangle_U = \sum_{x,y} y^2 \left| U_{x,y}^{(n)} \right|^2 = \frac{1}{4\pi^2} \int_{k_y=0}^{2\pi} \int_{k_x=0}^{2\pi} u_{k_x,k_y}^{(n)*} \left( i \frac{d}{dk_y} \right)^2 u_{k_x,k_y}^{(n)} dk_x dk_y$$

$$= \frac{n^2}{8\pi^2} \int_{k_y=0}^{2\pi} \int_{k_x=0}^{2\pi} \frac{\sin^2(k_x) \cos^2(k_y) \left( 1 + \sin^2(k_x) \cos(2k_y) \right)}{\left( 1 - \sin^2(k_x) \sin^2(k_y) \right)^2} dk_x dk_y$$

$$+ \frac{n^2}{8\pi^2} \int_{k_y=0}^{2\pi} \int_{k_x=0}^{2\pi} \frac{\sin^2(k_x) \cos^2(k_x) \cos^2(k_y)}{\left( 1 - \sin^2(k_x) \sin^2(k_y) \right)^2} \cos\left( 2n \arccos\left( \sin(k_x) \sin(k_y) \right) \right) dk_x dk_y$$

$$+ \frac{3n}{8\pi^2} \int_{k_y=0}^{2\pi} \int_{k_x=0}^{2\pi} \frac{\cos^4(k_x) \sin(k_x) \sin(k_y)}{\left( 1 - \sin^2(k_x) \sin^2(k_y) \right)^{5/2}} \sin\left( 2n \arccos\left( \sin(k_x) \sin(k_y) \right) \right) dk_x dk_y$$

$$- \frac{n}{4\pi^2} \int_{k_y=0}^{2\pi} \int_{k_x=0}^{2\pi} \frac{\cos^2(k_x) \sin(k_x) \sin(k_y)}{\left( 1 - \sin^2(k_x) \sin^2(k_y) \right)^{3/2}} \sin\left( 2n \arccos\left( \sin(k_x) \sin(k_y) \right) \right) dk_x dk_y$$

$$+ \frac{3}{4\pi^2} \int_{k_y=0}^{2\pi} \int_{k_x=0}^{2\pi} \frac{\sin^2(k_x) \cos^2(k_x) \cos^2(k_y)}{\left( 1 - \sin^2(k_x) \sin^2(k_y) \right)^3} \sin^2\left( n \arccos\left( \sin(k_x) \sin(k_y) \right) \right) dk_x dk_y$$

$$- \frac{1}{2\pi^2} \int_{k_y=0}^{2\pi} \int_{k_x=0}^{2\pi} \frac{\sin^2(k_x) \cos^2(k_x) \cos^2(k_y)}{\left( 1 - \sin^2(k_x) \sin^2(k_y) \right)^2} \sin^2\left( n \arccos\left( \sin(k_x) \sin(k_y) \right) \right) dk_x dk_y$$

In the limit of large $n$, we have:

$$P_U \to \frac{1}{2} + \frac{1}{8\pi^2} \int_{k_y=0}^{2\pi} \int_{k_x=0}^{2\pi} \frac{\cos^2(k_y) \sin^2(k_x)}{\sin^2\left( \arccos\left( \sin(k_x) \sin(k_y) \right) \right)} dk_x dk_y = 1 - \frac{1}{\pi}$$

$$\langle x \rangle_U \to 0$$

$$\langle y \rangle_U \to \left(1 - \frac{2}{\pi}\right) n$$

$$\langle x^2 \rangle_U \to \frac{n^2}{8\pi^2} \int_{k_y=0}^{2\pi} \int_{k_x=0}^{2\pi} \frac{\cos^2(k_x) \sin^2(k_y) \left(1 + \sin^2(k_x) \cos(2k_y)\right)}{\left(1 - \sin^2(k_x) \sin^2(k_y)\right)^2} dk_x dk_y = \left(1 - \frac{5}{2\pi}\right) n^2$$

$$\langle y^2 \rangle_U \to \frac{n^2}{8\pi^2} \int_{k_y=0}^{2\pi} \int_{k_x=0}^{2\pi} \frac{\sin^2(k_x) \cos^2(k_y) \left(1 + \sin^2(k_x) \cos(2k_y)\right)}{\left(1 - \sin^2(k_x) \sin^2(k_y)\right)^2} dk_x dk_y = \left(1 - \frac{13}{6\pi}\right) n^2$$

For the first few steps, we have:

$$u_{k_x,k_y}^{(0)} = 1 \to U_{x,y}^{(0)} = \delta(x) \delta(y)$$

$$u_{k_x,k_y}^{(1)} = i e^{-ik_y} \sin(k_x) \to U_{x,y}^{(1)} = \begin{cases} 0.5 \, (x=-1, y=1) \\ -0.5 \, (x=1, y=1) \\ 0 \, (Otherwise) \end{cases}$$

$$u_{k_x,k_y}^{(2)} = -e^{-ik_y} \left(\cos(k_y) + i \cos(2k_x) \sin(k_y)\right) \to U_{x,y}^{(2)} = \begin{cases} -0.25 \, (x=-2, y=0) \\ -0.5 \, (x=0, y=0) \\ -0.25 \, (x=2, y=0) \\ 0.25 \, (x=-2, y=2) \\ -0.5 \, (x=0, y=2) \\ 0.25 \, (x=2, y=2) \\ 0 \, (Otherwise) \end{cases}$$

$$u_{k_x,k_y}^{(3)} = i e^{-ik_y} \sin(k_x) \left(\cos(2k_x) \cos(2k_y) - 2\cos^2(k_x) + i \sin(2k_y)\right) \to U_{x,y}^{(3)} = \begin{cases} 0.125 \, (x=-3, y=-1) \\ 0.125 \, (x=-1, y=-1) \\ -0.125 \, (x=1, y=-1) \\ -0.125 \, (x=3, y=-1) \\ -0.25 \, (x=-3, y=1) \\ -0.25 \, (x=-1, y=1) \\ 0.25 \, (x=1, y=1) \\ 0.25 \, (i=3, y=1) \\ 0.125 \, (x=-3, y=3) \\ -0.375 \, (x=-1, y=3) \\ 0.375 \, (x=1, y=3) \\ -0.125 \, (x=3, y=3) \\ 0 \, (Otherwise) \end{cases}$$

These are consistent with what we expect from recursive analysis of evolution.

### B. Time independent phase modulation:

*Uniform modulation:*

For the case of time independent linear phase modulation, $\phi_{n,y} = y\phi$, we have:

$$\begin{bmatrix} s_{k_x,k_y}^{(n+1)} \\ p_{k_x,k_y}^{(n+1)} \end{bmatrix} = \begin{bmatrix} ie^{-ik_x}\sin(k_y) & e^{ik_x}\cos(k_y) \\ -e^{-ik_x}\cos(k_y) & -ie^{ik_x}\sin(k_y) \end{bmatrix} \begin{bmatrix} s_{k_x,k_y+\phi}^{(n)} \\ p_{k_x,k_y-\phi}^{(n)} \end{bmatrix}$$

For rational values of $\phi/2\pi = p/q$, we can consider $s_{k_x,k_y}$, $s_{k_x,k_y+\phi}$, $s_{k_x,k_y+2\phi}$, ..., $s_{k_x,k_y+(q-1)\phi}$ as well as $p_{k_x,k_y}$, $p_{k_x,k_y+\phi}$, $p_{k_x,k_y+2\phi}$, ..., $p_{k_x,k_y+(q-1)\phi}$ together as a vector. Then we can obtain the evolution equation for this vector. Note that this approach does not work for irrational values of $\phi/2\pi$. By representing this vector as

$$\psi^{(n)} = \begin{bmatrix} s_{k_x,k_y}^{(n)} & s_{k_x,k_y+\phi}^{(n)} & \cdots & s_{k_x,k_y+(q-1)\phi}^{(n)} & p_{k_x,k_y}^{(n)} & p_{k_x,k_y+\phi}^{(n)} & \cdots & p_{k_x,k_y+(q-1)\phi}^{(n)} \end{bmatrix}^T,$$

we can write the evolution equation as:

$$\psi^{(n+1)} = M(k_x, k_y)\,\psi^{(n)}$$

In this equation, $M$ is the evolution matrix and its size linearly increases by increasing $q$.

For a general propagation matrix, the eigen-energies can be calculated from the following characteristic equation:

$$\left| M - e^{iE}I \right| = 0$$

For instance, the evolution matrix for the case of $\phi = \pi$ is given by:

$$M = \begin{bmatrix} 0 & ie^{-ik_x}\sin(k_y) & 0 & e^{ik_x}\cos(k_y) \\ ie^{-ik_x}\sin(k_y+\pi) & 0 & e^{ik_x}\cos(k_y+\pi) & 0 \\ 0 & -e^{-ik_x}\cos(k_y) & 0 & -ie^{ik_x}\sin(k_y) \\ -e^{-ik_x}\cos(k_y+\pi) & 0 & -ie^{ik_x}\sin(k_y+\pi) & 0 \end{bmatrix}$$

In this case, the eigen-energies are given by:

$$E(k_x, k_y) = \pm \arccos\left(\pm\sqrt{1 - \sin^2(k_x)\sin^2(k_x)}\right)$$

Similar expressions can be obtained for other phase modulations. The obtained band diagrams for different values of $\phi$ are shown in Fig. S2.

Considering the evolution of the random walk under the effect of time independent gauge field, Fig. S3 summarizes the numerical results for the variation of quadratic means with time step for different values of $\phi$. The obtained results show that how such a gauge field can affect quadratic means of $x$ and $y$ to be decreased relative to the zero phase modulation case. Note that these quadratic means are normalized relative to the total power in the up and down channels.

*Non-uniform modulation:*

We next consider the case in which the phase modulation pattern changes across a boundary. This type of phase modulation can lead to the emergence of the edge states right along the boundary. Here, we have considered $y = 0$ as the boundary separating two time independent linear phase modulations in $y > 0$ and $y < 0$. Since we do not have translational symmetry in $y$ direction anymore, we cannot define $k_y$ for this configuration. However, we can still obtain the allowed eigen-energies for different $k_x$ values. In order to obtain these eigen-energies, we should obtain the

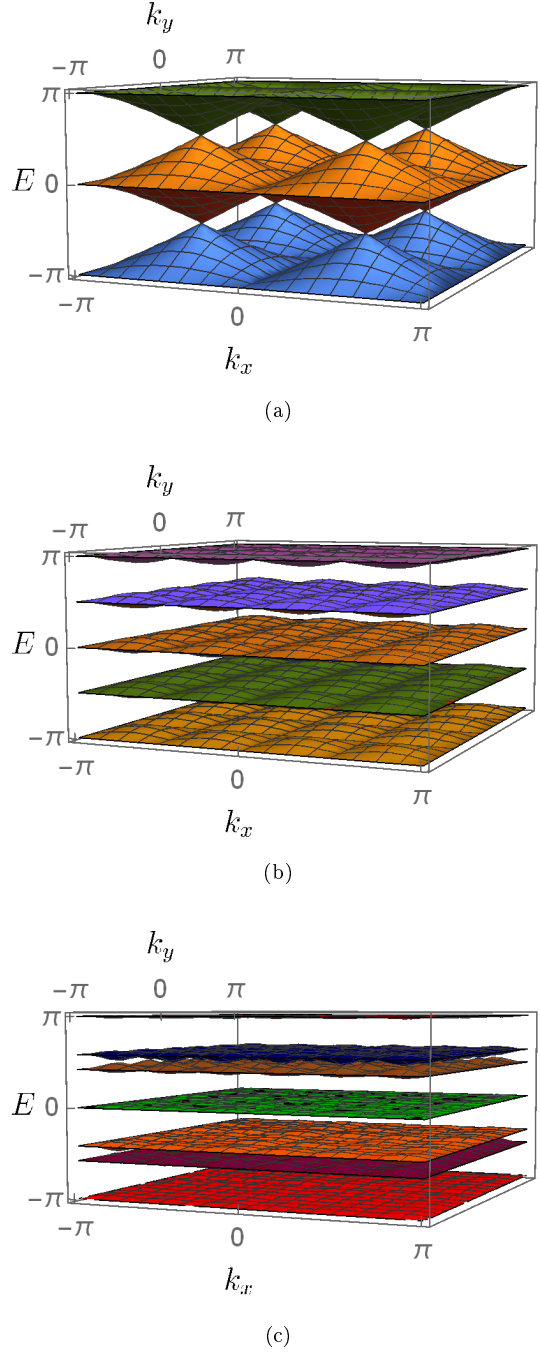

Figure S2. The energy band diagrams for the time independent phase modulation with (a) $\phi = \pi$ and (b) $\phi = \pi/2$ and (c) $\phi = \pi/3$.

propagation matrix in the real space under the appropriate boundary conditions. Depending on the phase modulations at the top ($\phi_T$) and the bottom ($\phi_B$), we obtain different band diagrams. Numerical results show that if one of $\phi_T$ and $\phi_B$ be zero then there exists no edge states. Moreover, the edge states exist only if $0 < \phi_T < \pi \,\&\, -\pi < \phi_B < 0$ or $0 < \phi_B < \pi \,\&\, -\pi < \phi_T < 0$.

We have plotted the energy band diagrams as functions of $k_x$ momentum for different $\phi_T$ and $\phi_B$ in Fig. S4. In this figure, we have considered different cases of $\phi_T = -\phi_B$ as non-uniform configurations with a boundary separating two different linear phase modulations. Note that swapping the values of $\phi_T$ with $\phi_B$ will modify the band diagram by replacing the edge states with the ones that propagate in the reverse direction. We have also shown the band

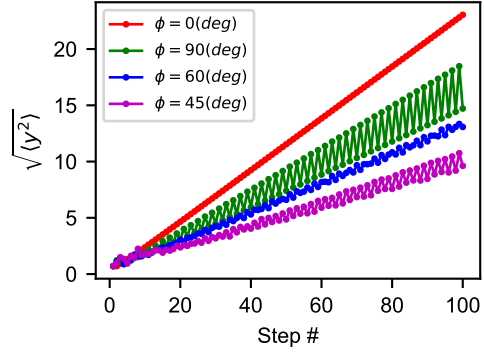

(a)

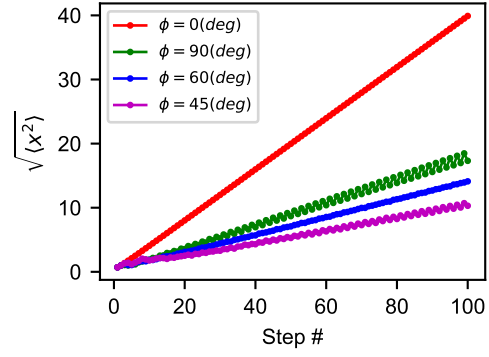

(b)

Figure S3. (a) The quadratic mean of $y$ and (b) the quadratic mean of $x$ as a function of time step for different values of phase modulations.

diagrams corresponding to uniform modulations with $\phi_B = \phi_T$ for comparison.

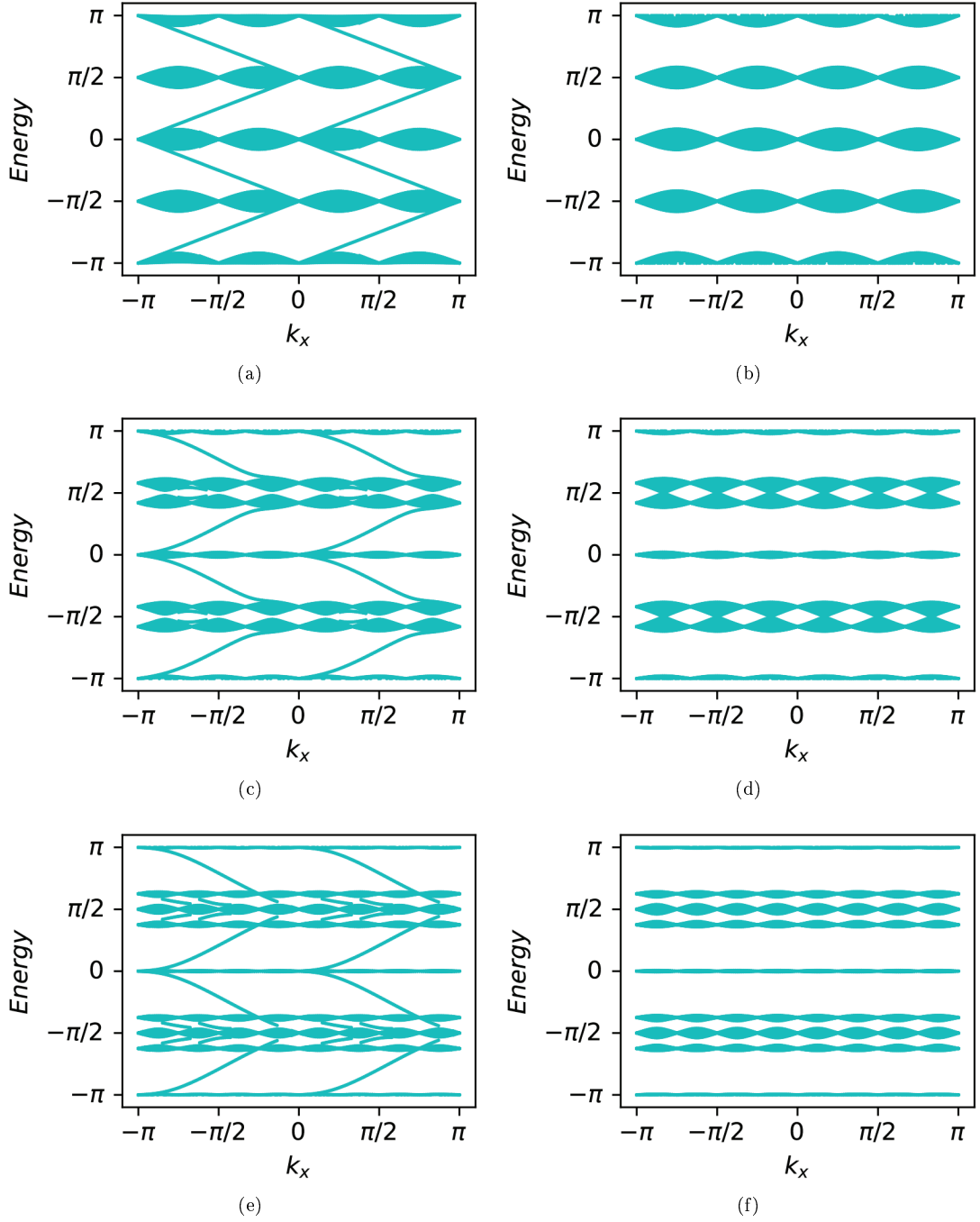

Figure S4. The left panel shows the energy band diagrams for the time independent non-uniform phase modulations of (a) $\phi_T = -\phi_B = \pi/2$ and (c) $\phi_T = -\phi_B = \pi/3$ and (e) $\phi_T = -\phi_B = \pi/4$. The right panel shows the corresponding band diagrams for the uniform phase modulations of (b) $\phi_T = \phi_B = \pi/2$ and (d) $\phi_T = \phi_B = \pi/3$ and (f) $\phi_T = \phi_B = \pi/4$.



\end{document}